\documentclass{kluwer}    
\usepackage[]{graphicx}
\begin{document}
\newcommand\Msun {M_{\odot}\ }

\begin{article}
\begin{opening}
\title{Dwarf Galaxies: Important Clues to Galaxy Formation}
\author{Eline \surname{Tolstoy}}
\runningauthor{Eline Tolstoy}
\runningtitle{Dwarf Galaxies}
\institute{Kapteyn Institute, University of Groningen, the Netherlands}
\date{\filedate}

\begin{abstract}
The smallest 
dwarf galaxies are the most straight forward objects in which to study
star formation processes on a galactic scale.  They are typically
single cell star forming entities, and as small potentials in orbit
around a much larger one they are unlikely to accrete much (if any)
extraneous matter during their lifetime (either intergalactic gas, or
galaxies) because they will typically lose the competition with the
much larger galaxy.  We can utilise observations of stars of a range
of ages to measure star formation and enrichment histories back to the
earliest epochs.  The most ancient objects we have ever observed in
the Universe are stars found in and around our Galaxy.  Their
proximity allows us to extract from their properties detailed
information about the time in the early Universe into which they were
born. A currently fashionable conjecture is that the earliest star
formation in the Universe occurred in the smallest dwarf galaxy sized
objects.

Here I will review some recent observational highlights in the study
of dwarf galaxies in the Local Group and the implications for
understanding galaxy formation and evolution.
\end{abstract}
\end{opening}           
\vskip -1cm
\section{Introduction}
One of the fundamental pillars of current structure formation models
(the CDM paradigm) is that small galaxies are the building blocks of
larger ones ({\it e.g.}, Navarro, Frenk \& White 1995).  Thus, the
dwarf galaxies around our Galaxy are arguably the remnants of the
formation of the Milky Way and as such provide a unique laboratory for
the detailed study of generic galactic assembly processes.  Whilst CDM
has been quite successful at modelling clusters of galaxies and
large-scale structure it currently faces problems on the small, dwarf
galaxy, scale.  It appears to over-predict the number and the mass
spectrum of satellites seen around galaxies such as our own (e.g.,
Moore {\it et al.}  1999) and there also appear to be inconsistencies
with regard to the timescale of the build up of larger galaxies (e.g.,
Prantzos \& Silk 1998), and the differences in the stellar populations
of large and small galaxies (e.g., Tolstoy {\it et al.} 2002).  These
problems might arise only because we have not yet made detailed enough
studies of our neighbours; there are still quite a number of
uncertainties in our interpretation of current observations. However,
results to date provide some fairly sizeable obstacles to the current
standard implementation of CDM on small scales.

\begin{figure} 
\begin{center}
\includegraphics[width=10cm]{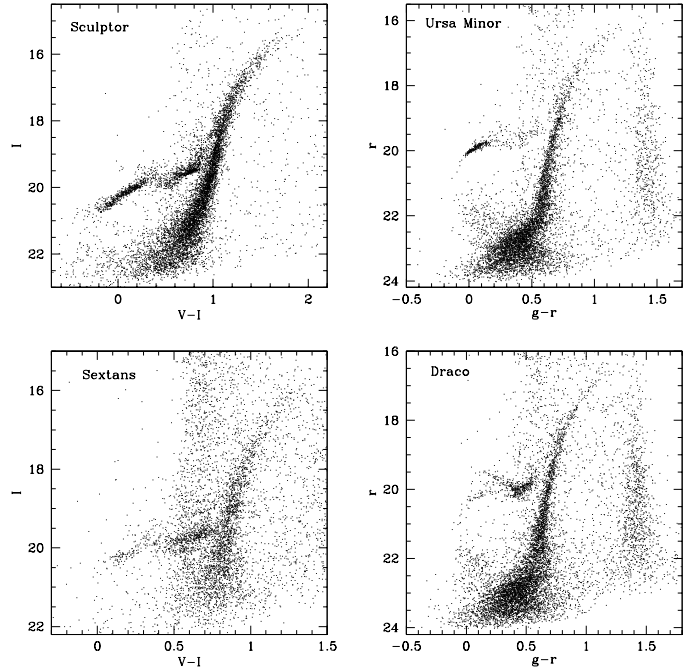}
\vskip -0.7cm
\caption{
Wide Field CMDs of Sculptor, Ursa Minor, Draco \& Sextans, taken with
the INT/WFC and the ESO2.2m/WFI covering 30$^\prime$ field of view for
each galaxy (Letarte, Irwin \& Tolstoy 2003, in prep.). These are all
old galaxies, but their CMDs look different in several crucial
respects. 
}
\label{wfcmds}
\end{center}
\end{figure}

\section{The Local Group Dwarf Population}
The majority of galaxies in the Local Group, by number, are Dwarf
Galaxies (35 out of 42 at the last count; 17 dSph, 5 dEs and 13 dIs,
e.g., van den Bergh 2000).  The definition of a dwarf galaxy is not a
fixed one (cf. Tammann 1993).  Half the galaxies in the Local Group
have a total mass, M$_{tot} \leq 3 \times 10^7\Msun$, and these fall
into two classes: dwarf Spheroidals (dSph) and dwarf Irregulars
(dI). These are the galaxies considered in this review.

\subsection{Dwarf Spheroidal Galaxies}
Dwarf spheroidal galaxies (dSph) are the smallest, faintest galaxies
we know of. Within the Local Group they are mostly satellites of
larger systems such as our Galaxy and M~31.  Some of these dSph formed
all their stars more than 10$-$12~Gyr ago and have apparently done
nothing at all since then, some have formed the majority of their
stars at intermediate times (6$-$8~Gyr ago), and a few have
experienced star formation as recently as 1$-$2~Gyr ago (see Mateo
1998, and references therein). All of the dSph galaxies, without
exception, have an ancient stellar population. Even the oldest and
simplest have had a complex star formation history (SFH), and these
systems, with virtually identical SFHs, still have different CMDs from
each other (see Figure~\ref{wfcmds}). Perhaps this is a result of
different environmental influences (e.g., Mayer
{\it et al.} 2001).  The evolution of dSph must be influenced, maybe
strongly, by the presence of our Galaxy.  The dynamical friction of
their orbits may have a strong (and varying) influence on the rate of
star formation (SFR).  No dSph around our Galaxy currently
has an (obviously) associated ISM.

\subsection{Dwarf Irregular Galaxies}
Dwarf Irregular (dI) galaxies have many similarities to dSph, but they
typically contain HI gas, often a large fraction
by mass, and also recent star formation. They all contain an
underlying old stellar population.  It is possible that they are dSph
witnessed in a more active state.  If a small dI stops forming stars
for a few 100~Myr it may look like a dSph. However, in general, dI
appear to have had a more constant, less disrupted, SFR over time and
have also typically attained higher metallicities than dSph.  Perhaps
dI, unlike dSph, are sufficiently massive to attain the threshold for
retaining their ISM during supernova explosions (e.g., Ferrara \&
Tolstoy 2000) or are not so disrupted by their 
environment, being typically more distant from larger systems, they
can sustain a more or less constant SFR, albeit at a low level,
over their history.  

\section{Observations of Nearby Dwarf Galaxies}
There are many different approaches to studying the evolutionary
history of a galaxy and the most reliable is to look directly at the
detailed properties of individual stars.  This is, however, restricted
to galaxies in the very nearby Universe. Although there is much to be
learnt from studies of integrated properties of more distant dwarf
galaxies ($>$2~Mpc away) I will not address this subject at all.

\begin{figure} 
\begin{center}
\includegraphics[width=10cm]{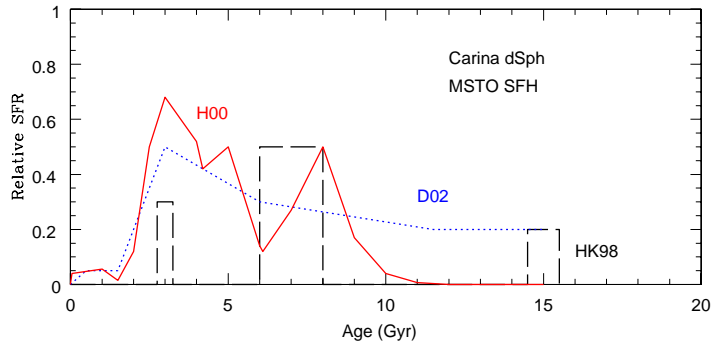}
\vskip -5.3cm
\caption{
Different SFHs for Carina dSph from the literature (arbitrarily
normalised).  Dashed lines from Hurley-Keller {\it et al.} (1998 =
HK98) from wide-field ground-based CTIO imaging.  The solid line
(Hernandez {\it et al.}  2000 = H00) and dotted line (Dolphin 2002 =
D02) use the same HST/WFPC2 dataset.  There are clear differences
between the three.  HK98, in contrast to H00 and D02 see discrete
bursts of star formation. This might be because HK98 are better able
to resolve such details because they cover a much larger area of the
galaxy and hence all the MSTOs are much better populated than in the
tiny HST field used by the other two studies.  Also D02 and HK98 find
evidence of ancient star-formation (as expected from the RR~Lyr
population detected by Saha, Seitzer \& Monet 1986), whereas H00 do
not.  The most critical difference is the assumed metallicity, not
only the absolute value, but the spread. D02 finds 
[Fe/H]=$-1.2\pm$0.4, where as HK98 assumes [Fe/H]=$-2.1\pm$0.1 and H00
[Fe/H]=$-2.0\pm$0.2.
}
\label{car}
\end{center}
\end{figure}

\subsection{Imaging}
The only way to measure the properties of a significant fraction of
the resolved stellar population in a galaxy is through multi-colour
imaging. The Hubble Space Telescope (HST) has produced significant
advances in this field over the last ten years (e.g., see Tolstoy 2000
and references therein). It has allowed the detailed analysis of
Colour-Magnitude Diagrams (CMDs) of resolved stellar populations to
determine SFHs out to distances around 5~Mpc.  However, the most
detailed studies come from galaxies within the Local Group at less
than 1~Mpc distance (see Tolstoy 2000, and references therein).

The dSph around the Milky Way are among the few galaxies in the
Universe for which we have accurate main sequence turnoff (MSTO) ages
going back to the epoch of earliest star formation (e.g., Carina:
Hurley-Keller, Mateo \& Nemec 1998; Leo I: Gallart {\it et al.}  1999;
Fornax: Buonanno {\it et al.}  1999). MSTOs are the most accurate
measurements of the age distribution of a stellar population and even
so there are problems in converting turnoff luminosities and colours
into accurate {\it absolute} ages, and thus CMDs into SFHs (see
Figure~\ref{car}).  This revolves around the well known and much
lamented age-metallicity degeneracy. It means that without {\it
independent} metallicity information it is not possible to uniquely
determine the age of a star from its colour and magnitude alone. A
major difference between the SFH determinations in Figure~\ref{car} is
the different metallicity evolution determined or assumed by each
study.

\subsection{Spectroscopy}
As telescopes are increasing in size and spectrographs are becoming
more sensitive and multiplexing capabilites are increasing it is
possible to take spectra of a significant number of individual stars
in nearby galaxies and determine the abundances of many different
elements and thus help to overcome the degeneracy inherent in
photometric measurements. FLAMES on the VLT with its 130 fibres over a
25$^\prime$ diameter field of view is eagerly awaited in this respect.

\subsubsection{Medium Resolution: The Ca~II Triplet}
The VLT instruments FORS1 and FORS2 in multi-object spectroscopy mode
are ideal for intermediate resolution spectroscopy of individual stars
in dwarf galaxies to determine metallicities (e.g., Tolstoy {\it et
al.}  2001) and radial velocities (e.g., Tolstoy \& Irwin 2000; Irwin
\& Tolstoy 2002) from the Ca~II triplet (CaT) lines for a significant
number of RGB stars.  Although the CaT provides a basic estimate of
[Fe/H] it was found to be broadly consistent with subsequent high
resolution UVES observations (Tolstoy {\it et al.} 2002). This is thus
a valuable method of obtaining [Fe/H] estimates at distances beyond
the limits of high resolution spectroscopy ($\geq$~250~kpc). This is
the only way to determine abundances of RGB stars of different ages in
dI galaxies, all of which are more distant than 450~kpc.

\subsubsection{High Resolution: Full Abundance Analysis}
With High Resolution Spectrographs such as UVES on the VLT, we can
observe individual stars in nearby dwarf galaxies and seek answers to
detailed questions about the enrichment {\it history} of a variety of
different elements within galaxies other than our own (e.g., Tolstoy
{\it et al.} 2002; Shetrone {\it et al} 2002).  Abundance patterns can
constrain the effects of the SFH on chemical evolution (e.g.,
McWilliam~1997).

A wealth of information is available in every high resolution
spectrum. The elemental abundances that can be measured fall into four
broad categories:

\noindent{\bf The Light Elements} ({\it e.g.}, O, Na, Mg, Al) allow us
to trace ``deep-mixing'' abundance patterns in RGB stars. This is a
very distinctive pattern that is markedly different in globular
cluster and field stars.

\noindent{\bf The $\alpha$-elements} ({\it e.g.}, O, Mg, Si, Ca, Ti),
the production of which is dominated by Type~II Supernovae.  The
$\alpha$-abundance limits the number that can have polluted the gas
from which the star was made. They also affect the age estimates based
on RGB isochrones, as lower $\alpha$-tracks are bluer than high
$\alpha$-tracks.

\noindent{\bf The Fe-Peak elements} ({\it e.g.}, V, Cr, Mn, Co, Ni,
Cu, Zn) are mostly believed to be the products of explosive
nucleosynthesis.  They can (in principle) limit the most massive
progenitor to have exploded in a galaxy ({\it e.g.}, Woosley \& Weaver
1995).

\noindent{\bf Heavy Metals} ({\it e.g.}, Y, Ba, Ce, Sm, Eu) enable a
distinction to be made between the fraction of s-process and r-process
elements in a star, and thus put detailed constraints on the number
and type of past Supernovae explosions.  The [Ba/Eu] ratio appears to
be an indicator of the contribution of AGB stars to the chemical
evolution process which provides yet another measure of the timescale
for chemical enrichment,

\begin{figure} 
\begin{center}
\includegraphics[width=10cm]{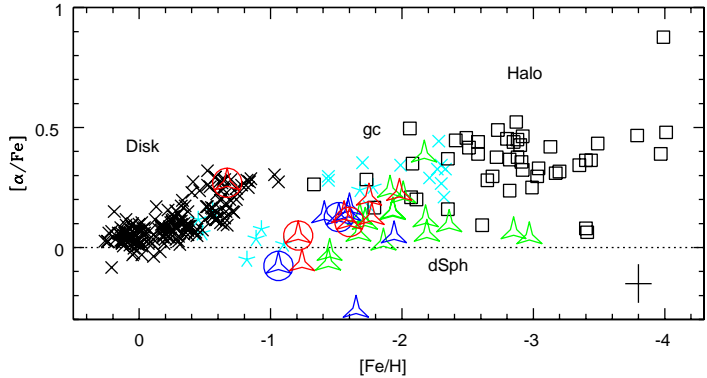}
\vskip -5.3cm
\caption{
The $\alpha$-abundances for dSph stars from Tolstoy {\it et al.}
(2002) and Shetrone {\it et al.} (2001) plotted versus [Fe/H].  The
triangle and circle symbols are the individual stars observed in
Carina, Leo~I, Sculptor, Fornax, Draco, Ursa Minor and Sextans (see
Tolstoy {\it et al.} for more details). The crosses (at [Fe/H]$< -1$)
are Galactic disk star measurements from Edvardsson {\it et al.} 1993;
the open squares are halo data from McWilliam {\it et al.} 1995 and
the stars are UVES data from a study of LMC star clusters of different
ages from Hill {\it et al.} 2000.  Crosses (at [Fe/H] $> -1$) are
Galactic globular cluster measurements (gc).  A representative error
bar is also plotted.  This plot highlights the differences between the
$\alpha$-element abundances of stars observed in different
environments.
}
\label{alf1}
\end{center}
\end{figure}

\section{Understanding the Chemical Evolution of Galaxies}
A sample of 15 RGB stars were observed in 4 southern dSph (Sculptor,
Fornax, Carina and Leo~I) with VLT/UVES (Shetrone {\it et al.} 2002;
Tolstoy {\it et al.} 2002), and 17 RGB stars were observed in 3
northern dSph (Draco, Ursa Minor, Sextans) with Keck/HIRES (Shetrone,
Bolte \& Stetson 1998; Shetrone, C\^{o}t\'{e} \& Sargent 2001).
Combining these surveys gives us detailed abundances for individual
stars in dSph around our Galaxy covering a range of SFHs.  There have
also been detailed abundance studies in LMC star clusters (e.g., Hill
{\it et al.} 2000) and also in the disk of our Galaxy (Edvardsson {\it
et al.} 1993) and in our halo (McWilliam {\it et al.}  1995) which
allow us to compare dSph with other environments (e.g., the
$\alpha$-elements, see Figure~\ref{alf1}).

\begin{figure} 
\begin{center}
\includegraphics[width=10cm]{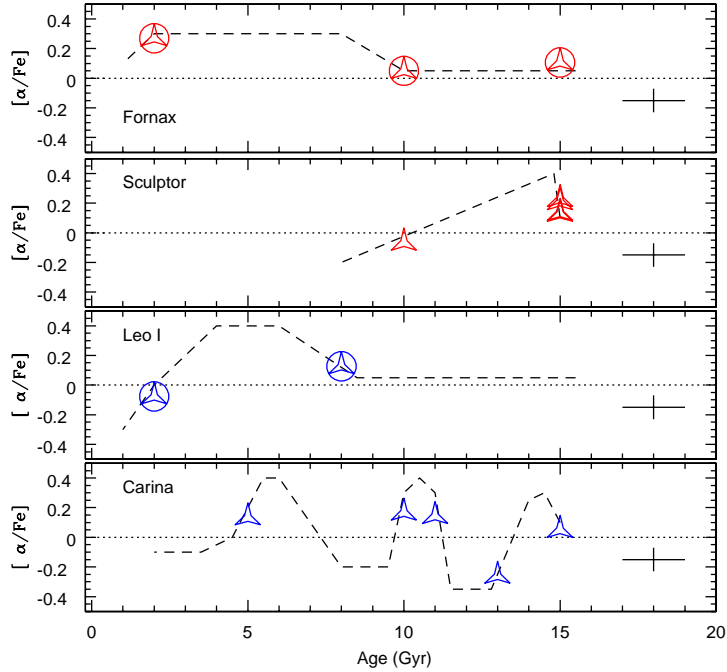}
\vskip-0.5cm
\caption{
An {\it illustrative} scenario which might allow us to tie in our
determinations of star formation history with [$\alpha$/Fe] for each
galaxy (see Tolstoy {\it et al.} 2002 for more details and caveats).
The symbols are defined as in Figure~3.  Representative error bars are
plotted. It is obvious that the dashed lines cannot be constrained
with the few data we have.  
}
\label{alf2}
\end{center}
\end{figure}

The $\alpha$-abundances can be plotted both in the ``traditional''
manner, against [Fe/H] (see Figure~\ref{alf1}) and against age (see
Figure~\ref{alf2}). Both provide differing insights as to how galaxies
are evolving with time, and also how observations of stars in dSph
compare to those in our Galaxy and in the Magellanic Clouds. Plotting
against age is more useful from the point of view of understanding
chemical evolution, but it is not easy to find suitable measurements
with which to compare dSph results as it is challenging to determine
accurate ages for stars in our Galaxy. It is somewhat more straight
forward in the simpler environment of dSph, although care is still
required.

The dSph $\alpha$-abundances when plotted against age appear to follow
the same distribution as those for our disk and the LMC star clusters
(see Tolstoy {\it et al.} 2002).  However, if we look at the plot of
$\alpha$ versus [Fe/H] in Figure~\ref{alf1}, the properties of dSph
are significantly different from the disk in the sense that although
the levels and the variation with stellar age of the $\alpha$-elements
are similar this is occurring at significantly lower [Fe/H] in the
dSph.  Figure~\ref{alf1} shows that the properties of the dSph differ
from those typical of halo stars, although there is overlap, such that
the [$\alpha$/Fe] of the halo stars are typically higher.  It is as if
all stars know the mass of the potential in which they are forming.

\subsection{Interpretation of Abundances}
The low [$\alpha$/Fe] found in the disk stars of our Galaxy has been
interpreted as evidence for star-formation in material with a large
fraction of Supernovae Ia ejecta ({\it e.g.}, Tinsley 1979; Gilmore \&
Wyse 1991).  This is perhaps not surprising for our disk, with high
metallicity, and typical predictions of fairly recent formation (from
pre-enriched material).  It is not clear that the same assessment can
be made of the similarly low [$\alpha$/Fe] for stars in dSph galaxies.
The same low [$\alpha$/Fe] is also found in the oldest stars, which is
at odds with the Supernovae Ia time scale.  This might be a remnant
of the initial enrichment of the dSph gas in the early universe, by a
process quite different from the star formation we see today.

Everything we know about dwarf galaxies suggests that they have never
had very high SFRs. The stars in dSph typically have much lower [Fe/H]
and [O/H] than in our disk. The low SFR means that Supernovae II
products may predominately come from low mass (8$-12 \Msun$)
progenitors, which result in lower [$\alpha$/Fe] than their higher
mass cousins ({\it e.g.}, Woosley \& Weaver 1995).  This is 
(unfortunately) effectively a
truncated IMF, but it is motivated by the likelihood that in the 
physical conditions to be found in 
small galaxies the probability of forming high
mass molecular clouds (and thus high mass stars) is low.

With the recent results of Tolstoy {\it et al.} (2002) and Shetrone
{\it et al.} (2001) it is for the first time possible to directly {\it
measure} the [$\alpha$/Fe] evolution (as well as other elements) of
the stellar populations of dSph over Gyr time scales, back from the
earliest epoch of formation to the most recent star formation.  The
range of variation in [$\alpha$/Fe] is quite small (which means
accurate measurements are required to observe it), and it never
reaches the parameter space where the disk and halo are stars
predominantly to be found. So probably star formation, when it occurs,
always occurs at similarly low levels in these small galaxies.

In Figure~\ref{alf2} the [$\alpha$/Fe] vs. age for stars in four dSph
galaxies is plotted separately, and over-plotted is an illustrative
{\it estimate} of the variation of [$\alpha$/Fe] for each galaxy {\it
given the star formation rate variation}. There really are not
sufficient data on these galaxies to be certain that we are seeing
direct evidence of evolution in [$\alpha$/Fe], but the results are
highly suggestive. The dashed lines are not derived from the SFH
directly, but a knowledge of the SFH is used to find the most likely
pattern with time in [$\alpha$/Fe].  Carina, for example, has the most
impressive evidence for evolution of abundances due to variations in
star formation rates. The variations seen in the $\alpha$ abundances
are supported by consistent variations in Ba, La, Nd and Eu (see
Shetrone {\it et al.} 2002).  However, more data are needed to confirm
these speculations.

\subsection{The Bottom Line for CDM}
The most recent VLT/UVES results (Tolstoy {\it et al.} 2002) combined
with Keck/HIRES data (Shetrone {\it et al.} 2001) unequivocally show
that the stars observed in dSph galaxies {\it today} (many of which
are extremely old) {\it cannot} be used to make up a significant
fraction of the stellar mass in our Galaxy, neither in the disk nor in
the inner-halo (nor the bulge) because their nucleosynthetic
signatures are not compatible. This places a limit on the time
(redshift) at which the majority of merging of small halos must have
occurred to create the Milky Way, if this is indeed to be the
formation mechanism.  This is of course assuming that the initial
small halos are similar to the dwarf galaxies we see today. These
recent abundance measurements require that the majority of these kinds
of mergers must have occurred very early, in the first few Gyr of
structure formation, because this is the only way to ensure that the
majority of star formation will occur in a deep potential with the
requisite conditions for massive star formation to explain the
abundance patterns seen in our Galaxy, but not in dSph (i.e. mergers
will add mostly gas to the larger system, but few stars).  More data
is needed to put these initial results on a firm statistical basis,
and of course dSph results do not place any limits on the effect of
significantly larger accretions, ({\it e.g.}, LMC like objects).
However, there are suggestions that the abundance patterns of stars in
the Clouds and other nearby Irregulars do not resemble our Galaxy
anymore than the dSph do ({\it e.g.}, Hill {\it et al.} 2000; Venn
{\it et al.} 2002, in prep).

The only component of our Galaxy which could plausibly contain a
significant contribution from stars formed in accreted dwarf galaxies
is the halo (e.g., Nissen \& Schuster 1997), and it contains only
about 1\% of the stellar mass of our Galaxy (e.g., Morrison 1993), and
only a fraction of this, the outer-halo ($\sim$~10\%), could plausibly
include stars accreted from dwarf galaxies (e.g., Unavane, Wyse \&
Gilmore 1996).

\vskip -0.3cm
\acknowledgements
I thank my collaborators on recent UVES programmes for introducing me
to the intricacies of high resolution stellar spectroscopy and for
their valuable insights into the subject of this review: Vanessa Hill,
Francesca Primas, Kim Venn \& Matt Shetrone.  Thanks also to Evan
Skillman and Sally Oey for useful conversations.
I thank the organisers of this conference for inviting
me.  I gratefully acknowledge a fellowship of the Royal Netherlands
Academy of Arts and Sciences.

\end{article}

\begin{thebibliography}{}

\bibitem[\protect\citeauthoryear{}{}]{}
Buonanno R., Corsi C.E., Castellani M., Marconi G. {\it et al.}
1999 \newblock {\em AJ}, 118, 1671


\bibitem[\protect\citeauthoryear{}{}]{}
Dolphin A. 
2002 \newblock {\em MNRAS}, 332, 91

\bibitem[\protect\citeauthoryear{}{}]{}
Edvardsson B., Andersen J., Gustafsson B.
{\it et al.} 
1993 \newblock{\em A\&A}, 275, 101

\bibitem[\protect\citeauthoryear{}{}]{}
Ferrara A. \& Tolstoy E. 
2000 \newblock{\em MNRAS}, 313, 291


\bibitem[\protect\citeauthoryear{}{}]{}
Gallart C, Freedman W.L., Aparicio A., Bertelli G. \& Chiosi C. 
1999 \newblock {\em AJ}, 118, 2245

\bibitem[\protect\citeauthoryear{}{}]{}
Gilmore G. \& Wyse R.F.G. 
1991 \newblock {\em ApJL}, 367, 55

\bibitem[\protect\citeauthoryear{}{}]{}
Hernandez X., Gilmore G. \& Valls-Gabaud D. 
2000 \newblock {\em MNRAS}, 317, 831

\bibitem[\protect\citeauthoryear{}{}]{}
Hill V., François P., Spite M., Primas F. \& Spite F. 
2000 \newblock {\em A\&A}, 364, L19


\bibitem[\protect\citeauthoryear{}{}]{}
Hurley-Keller D., Mateo M. \& Nemec J. 
1998, \newblock {\em AJ}, 115, 1840

\bibitem[\protect\citeauthoryear{}{}]{}
Irwin M.J. \& Tolstoy E.
2002, \newblock {\em MNRAS}, 336, 643


\bibitem[\protect\citeauthoryear{}{}]{}
Mateo M.
1998 \newblock {\em ARAA}, 36, 435 

\bibitem[\protect\citeauthoryear{}{}]{}
Mayer L., Governato F., Colpi M., Moore B., Quinn T. {\it et al.}
2001 \newblock {\em ApJL}, 547, 123

\bibitem[\protect\citeauthoryear{}{}]{}
McWilliam A. 
1997 \newblock {\em ARA\&A}, 35, 503

\bibitem[\protect\citeauthoryear{}{}]{}
McWilliam A., Preston G.W., Sneden C. \& Searle L. 
1995 \newblock {\em AJ}, 109, 2757

\bibitem[\protect\citeauthoryear{}{}]{}
Moore B., Ghigna S., Governato F., Lake G. 
{\it et al.} 
1999 \newblock {\em ApJL}, 524, 19

\bibitem[\protect\citeauthoryear{}{}]{}
Morrison H.L. 
1993 \newblock {\em AJ}, 106, 578

\bibitem[\protect\citeauthoryear{}{}]{}
Navarro J.F., Frenk C.S. \& White S.D.M. 
1995 \newblock {\em MNRAS},  275, 56

\bibitem[\protect\citeauthoryear{}{}]{}
Nissen P.E. \& Schuster W.J. 
1997 \newblock {\em A\&A}, 326, 751

\bibitem[\protect\citeauthoryear{}{}]{}
Prantzos N. \& Silk J.
1998 \newblock {\em ApJ}, 507, 229

\bibitem[\protect\citeauthoryear{}{}]{}
Saha A., Seitzer P., \& Monet D.G.
1986 \newblock {\em AJ}, 92, 302

\bibitem[\protect\citeauthoryear{}{}]{}
Shetrone M.D., Bolte M. \& Stetson P.B. 
1998 \newblock {\em AJ}, 115, 1888

\bibitem[\protect\citeauthoryear{}{}]{}
Shetrone M.D., C\^{o}t\'{e} P. \& Sargent W.L.W. 
2001 \newblock {\em ApJ}, 548, 592

\bibitem[\protect\citeauthoryear{}{}]{}
Shetrone M.D., Venn K.A., Tolstoy E., 
Primas F. 
{\it et al.}
2002 \newblock {\em AJ}, submitted 

\bibitem[\protect\citeauthoryear{}{}]{}
Tammann G.A. 
1993 \newblock {\em Dwarf Galaxies}, ed. Meylan \& Prugniel, p. 3

\bibitem[\protect\citeauthoryear{}{}]{}
Tinsley B.M. 
1979 \newblock {\em ApJ}, 229, 1046


\bibitem[\protect\citeauthoryear{}{}]{}
Tolstoy E.
2000 \newblock {\em A Decade of HST Science}, eds. Livio {\it et al.},
in press

\bibitem[\protect\citeauthoryear{}{}]{}
Tolstoy E. \& Irwin M.J. 
2000 \newblock {\em MNRAS}, 318, 1241


\bibitem[\protect\citeauthoryear{}{}]{}
Tolstoy E., Irwin M.J., Cole A.A., Pasquini L.
{\it et al.}
2001 \newblock {\em MNRAS}, 327, 918 

\bibitem[\protect\citeauthoryear{}{}]{}
Tolstoy E., Venn K.A., Shetrone M., Primas F.
{\it et al.}
2002 \newblock {\em AJ}, submitted

\bibitem[\protect\citeauthoryear{}{}]{}
Unavane M., Wyse R.F.G. \& Gilmore G. 
1996 \newblock {\em MNRAS}, 278, 727


\bibitem[\protect\citeauthoryear{}{}]{}
van den Bergh S.
2000 \newblock {\em PASP}, 112, 529

\bibitem[\protect\citeauthoryear{}{}]{}
Woosley S. E. \& Weaver T. A. 
1995 \newblock {\em ApJS}, 101, 181

\end{thebibliography}
\end{document}